\documentclass[fleqn,usenatbib]{mnras}
\usepackage[utf8]{inputenc}
\usepackage[T1]{fontenc}
\usepackage{graphicx}
\usepackage{amsmath}
\usepackage{amssymb}

\title[Simulations of the radio polarization of  a precessing pulsar PSR J1906+0746]
{Simulations of the radio polarization of  a precessing pulsar PSR J1906+0746}
\author[A.K. Galishnikova, A.A. Philippov and V.S. Beskin]
{A.K. Galishnikova,$^{1,2}$\thanks{E-mail:     \href{emailto:alisag@princeton.edu}{alisag@princeton.edu}}
A.A. Philippov$^{3,2}$ 
and V.S. Beskin$^{2,4}$
\\
$^1$Department of Astrophysical Sciences, Peyton Hall, Princeton University, Princeton, NJ 08544, USA\\
$^2$Moscow Institute of Physics and Technology, Dolgoprudny, Institutsky per., 9, Moscow Region, 141701, Russia\\
$^3$Center for Computational Astrophysics, Flatiron Institute, 162 Fifth Avenue, New York, NY 10010, USA\\
$^4$P.N. Lebedev Physical Institute, Leninsky Prosp., 53, Moscow 119991, Russia
}

\begin{document}
\label{firstpage}
\pagerange{\pageref{firstpage}--\pageref{lastpage}}
\maketitle

\date{Accepted, Received}

\begin{abstract}
The recently constructed theory of radio wave propagation in the pulsar magnetosphere outlines the general aspects of the radio light curve and polarization formation. It allows us to describe general properties of mean profiles, such as the position angle ($PA$) of the linear polarization, and the circular polarization for the realistic structure of the pair creation region in the pulsar magnetosphere. In this work, we present an application of the radio wave propagation theory to the radio observations of pulsar PSR J1906+0746. This pulsar is particularly interesting because observations of relativistic spin-precession in a binary system allows us to put strong constraints on its geometry. Because it is an almost orthogonal rotator, the pulsar allows us to observe both magnetic poles; as we show, this is crucial for testing the theory of radio wave propagation and obtaining constraints on the parameters of magnetospheric plasma. Our results show that plasma parameters are qualitatively consistent with theories of pair plasma production in polar cap discharges. Specifically, for PSR J1906+0746, we constrain the plasma multiplicity $\lambda \sim 10^3$ and the Lorentz-factor of secondary plasma $\gamma \sim $ a few hundred.\\
\end{abstract}

\begin{keywords}
Neutron stars -- radio pulsars -- polarization
\end{keywords}

\section{Introduction}
\label{sec:intro}

Radio pulsars were discovered more than fifty years ago~\citep{Hewish}; however, their radio emission mechanism still remains unknown~\citep{L&GS, L&K}. A large number of observational light curves (also called mean profiles), which include intensity, linear polarization and circular polarization, are available now~\citep{H&R, W&J, J&K}. The rotating vector model (hereafter abbreviated RVM,~\citealt{R&Cooke}) is the most frequently used theory to describe mean profiles of radio pulsars. In particular, it is usually used to infer the inclination angle $\alpha$ between magnetic and rotation axes and the impact angle $\beta$, which characterizes the closest approach between the observer direction and the magnetic axis ~\citep{L&M, Rankin, T&M}. However, full understanding of available pulsar radio profiles cannot be reached within this model. 

For instance, according to the RVM (more exactly, the ``hollow-cone'' model~\citealt{O&S}), the polarization characteristics of the observable radiation are formed not so far from the stellar surface, at radii $\leq 30R$ (hereafter, $R$ denotes the stellar radius). This assumption leads to almost exactly linear polarization of propagating electromagnetic waves due to the super-strong magnetic field in the region near the stellar surface. Thus, the RVM predicts fully linearly polarized radio emission, while observations show the presence of a substantial degree of circular polarization. Therefore, propagation effects have to be taken into account. As the wave propagates further outwards in the magnetosphere, the strength of the magnetic field drops, so the wave does not have to be fully linearly polarized.

Thus, propagation effects in the pulsar magnetosphere should play an important role in understanding observed characteristics of the pulsar radio emission. In particular, the interaction of electromagnetic waves with magnetospheric plasma must result in considerable deviation from the predictions of the RVM. Individually, all the main propagation effects, namely: magnetospheric plasma birefringence~\citep{B&A, BGI88, P&L1, P&L2}, cyclotron absorption~\citep{mikhailovskii82, abs1, luomelrose2004}, and limiting polarization~\citep{C&R, lim1, P&L2, lim3, andrianovbeskin2010, wanglaihan2010} had already been taken into account. But only recently have all these effects been successfully brought together (also taking into account the real structure of the magnetic field in the vicinity of the light cylinder where the dipolar  approximation is no longer applicable) ~\citep{PaperI}. Thus, it has become possible to quantitatively analyze the average profiles of pulsar radio emission~\citep{PaperII, Hakobyan}.

In this paper, we analyze recent observations obtained for radio pulsar PSR J1906+0746~\citep{Desvignes}, which has been observed for 13 consecutive years. The uniqueness of this pulsar, as along with PSR J0737-3039~\citep{Perera} and J1141-6545~\citep{Krishnan}, is that it is located in a close binary system. Misalignment of the spin vector of the pulsar with respect to the total angular momentum vector results in relativistic precession of its rotational axis. This leads to a change in viewing geometry and to observed significant changes in polarization profiles. In particular, observations show a simultaneous switching of the signs of the circular polarization $V$ and the derivative ${\rm d}PA/{\rm d}\phi$ at the moment when the line of sight crosses the magnetic axis (hereafter, $\phi$ denotes the pulsar rotational phase). As we describe below, this is broadly consistent with the prediction of the radio propagation theory described by~\citet{PaperI}. 

On the other hand, the consistent analysis of an almost orthogonal pulsar requires a substantial refinement of the theory. This is because previous work assumed the distribution of  the number density of secondary plasma to be symmetric about the magnetic axis. However, this is not a very good approximation for polar caps of nearly orthogonal pulsars~\citep{BaiSpitkovsky, TimokhinArons}. Below, following~\citet{TimokhinArons, Gralla, Lockhart}, we assume that the structure of the pair production regions is determined by the distribution of the magnetospheric current in the force-free magnetosphere. In this approach, which is valid if the size of  the particle acceleration region is significantly smaller than the polar cap radius,  pair creation is possible if the value of the longitudinal electric current density, \mbox{$j_{\parallel} = {\bf j \cdot B}/B$}, is larger than the Goldreich-Julian current{\footnote{The GJ charge density, $\rho_{\rm GJ} = - {\bf \Omega \cdot B}/(2\pi c)$, characterizes the minimal charge density required for the screening of the longitudinal electrical field in the pulsar magnetosphere.}} (hereafter referred to as the GJ current,~\citealt{GoldreichJulian})
\begin{equation}
j_{\rm GJ} = -\frac{{\bf \Omega \cdot B}}{2\pi},
\label{GJ}
\end{equation}
or if it has the opposite sign~\citep{TimokhinArons}. In order to identify such regions, we use analytic expressions for the 4-current in the magnetosphere of a neutron star~\citep{Gralla}. These formulas were obtained by fitting the results of force-free simulations and allow us to find regions which can support active pair production at the polar cap. We assume that observed radio emission originates from the region of active pair production. Therefore, our new numerical method includes the improved geometry of the emission region, as well as previously described propagation effects.

It is necessary to stress that the above criteria for the activity of pair production has been formulated in the one-dimensional case, when the size $l_{\rm gap}$ of the particle acceleration zone is significantly smaller than the size of the pulsar polar cap, $R_0$. According to~\citet{TimokhinHarding2015}, the size of the acceleration zone can be estimated as follows:
\begin{equation}
    l_{\rm gap} \simeq 2 \times 10^4 \chi_{\rm a}^{1/7} \xi_{j}^{-3/7} \rho_{\rm c,7}^{2/7}P^{3/7}B_{12}^{-4/7} {\rm cm},
\end{equation}
where $\chi_{\rm a}$ denotes the
value of the $\chi$ parameter (photon energy multiplied by the perpendicular magnetic
field) when the optical depth reaches 1, $\xi_j$ denotes the electric field, $P$ is the period in seconds, $\rho_{\rm c,7}=\rho_{\rm c}/10^7$ cm is the normalized curvature radius of the magnetic field lines and $B_{12}=B/10^{12}$ G is the normalized magnetic field. In particular, for PSR J1906+0746 ($P = 0.14$ s, ${\dot P}_{-15} = 20$), $l_{\rm gap} \sim 10^3$ cm, which is an order of magnitude less than the size of the polar cap \mbox{$R_{0} \simeq R (\Omega R/c)^{1/2} \sim 10^4$ cm.} Moreover, for an almost orthogonal pulsar, the polar cap current is super GJ, $j \gg j_{GJ}$, which strengthens the applicability of a 1D approximation.

In this paper, we simulate the mean radio profile of a nearly orthogonal rotator (i.e., we determine the intensity $I(\phi)$, the position angle curve $PA(\phi)$ of the linear polarization, and the circular polarization $V(\phi)$ along the mean profile), which is necessary to explain the observable evolution of the polarization characteristics of the precessing pulsar PSR J1906+0746. In Section~\ref{sec:obs} we discuss observational data presented by~\citet{Desvignes}. We summarize the main results of the propagation theory with a focus on an orthogonal rotator, present the results of our numerical simulations and compare these results with observation data in Section~\ref{sec:t&s}. Finally, we discuss our main findings in Section~\ref{sec:conslusion}.

\section{Observations}\label{sec:obs}

\begin{figure}
    \includegraphics[width=\columnwidth]{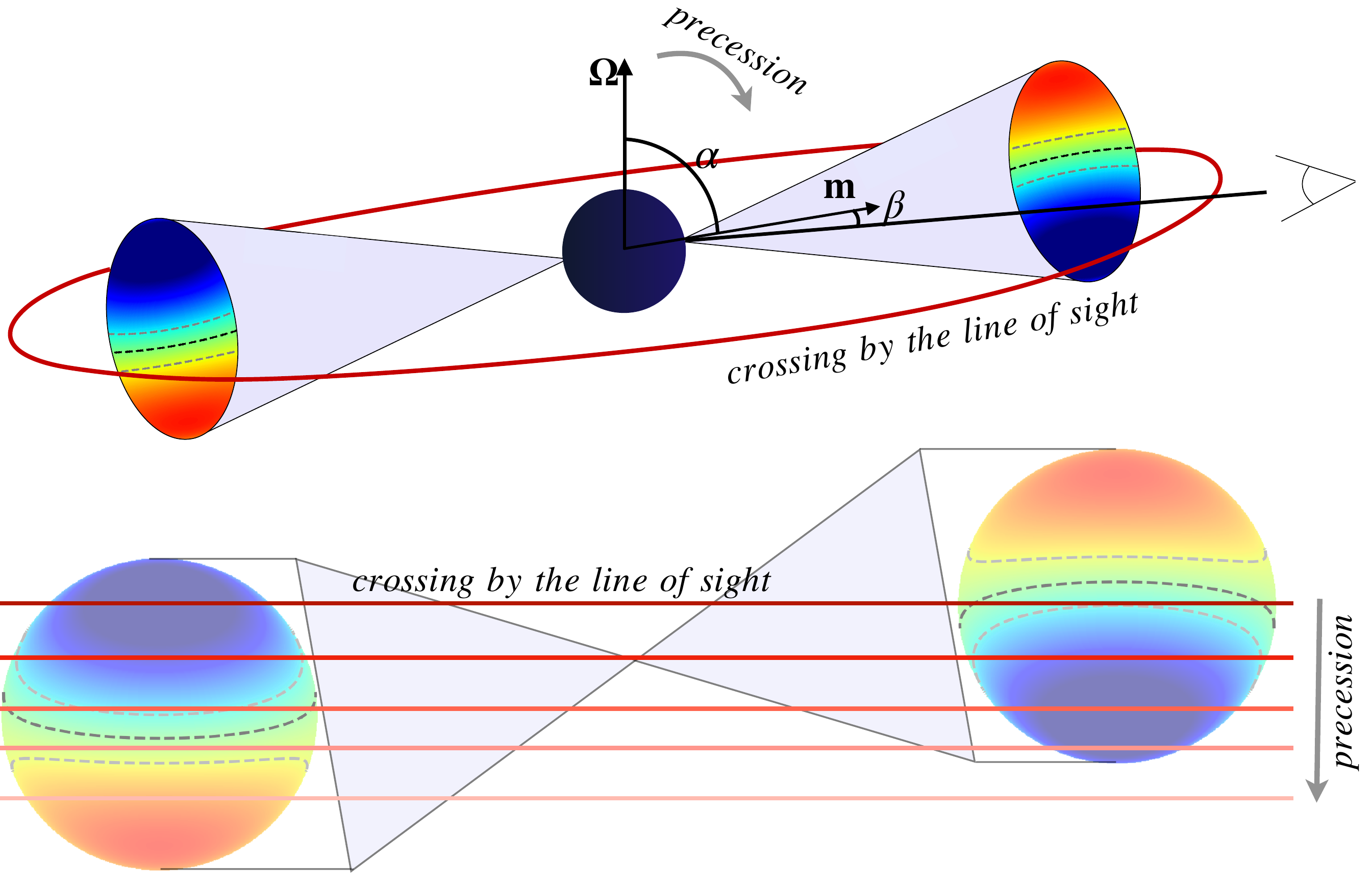}
    \caption{Illustration of the emission pattern of an almost orthogonal pulsar ($\alpha \sim 90^{\circ}$ is the inclination angle between the magnetic moment $\bf{\mu}$ and axis of rotation $\bf{\Omega}$). The observer's line of sight is directed at an angle $\beta$ with respect to the magnetic moment. Color represents the magnetospheric current density, $\xi = j_{\parallel}/j_{GJ}$; large positive (red) and negative (blue) values correspond to the regions that support active pair formation (see Section~\ref{sec:cap}). {\bf Top panel:} As the pulsar rotates, the observer's line of sight can cross the directivity pattern of both northern and southern polar caps (red solid line). Therefore, emission from both polar caps can be observed. {\bf Bottom panel:} a schematic illustration of the figure in the top panel. For certain values of $\beta$, emission from both polar caps can be observed. Moreover, if the pulsar precesses, the value of $\beta$ changes with time. Therefore, the relative position of the line of sight is evolving with time (gray line), and emission from different regions of both polar caps can be observed.}
    \label{fig:crossing}
\end{figure}

Pulsar PSR J1906+0746 was first discovered by the Parkes Multibeam Pulsar Survey in 1998~\citep{PMPS}, when only a single pulse was detected. A second peak at the phase $\phi = 180^{\circ}$ appeared in 2005~\citep{Lorimer}. This pulsar has since been observed for 13 years at 1.38 GHz with the 305-m William E. Gordon Arecibo radio telescope and the Puerto Rico Ultimate Pulsar Processing Instrument (PUPPI). The presence of an interpulse at the phase $\phi = 180^{\circ}$ indicates that this pulsar is almost orthogonal (the inclination angle between the magnetic moment and the rotational axis $\alpha \sim 90^{\circ}$). As one can see in Figure~\ref{fig:crossing}, this geometry allows us to observe the emission from both polar caps. In particular, due to the symmetry of the emission from both polar caps, the light curves for both pulses will be similar for an observer located exactly at the equator. As we have already emphasized, in the binary system containing pulsar PSR J1906+0746, the curvature of space-time near massive bodies leads to relativistic spin-orbit coupling. Therefore, the spins of those bodies precess around the total angular momentum, thus changing their relative orientation to an observer. According to~\citet{Desvignes}, from 2001 to 2018, the angle $\beta$ of the closest approach between the observer direction and the magnetic axis has changed from $5^{\circ}$ to $-22^{\circ}$ for the main pulse and from $20^{\circ}$ to $-6^{\circ}$ for the interpulse\footnote{These values of $\beta$ have been obtained by fitting an RVM curve into the position angle curve at different observational epochs.}. In other words, observations of PSR J1906+0746 have been able to scan a significant fraction of the emission region in both polar caps due to a gradual change in the angle $\beta$.  

\begin{figure*}
    \centering
    \includegraphics[width=0.9\textwidth]{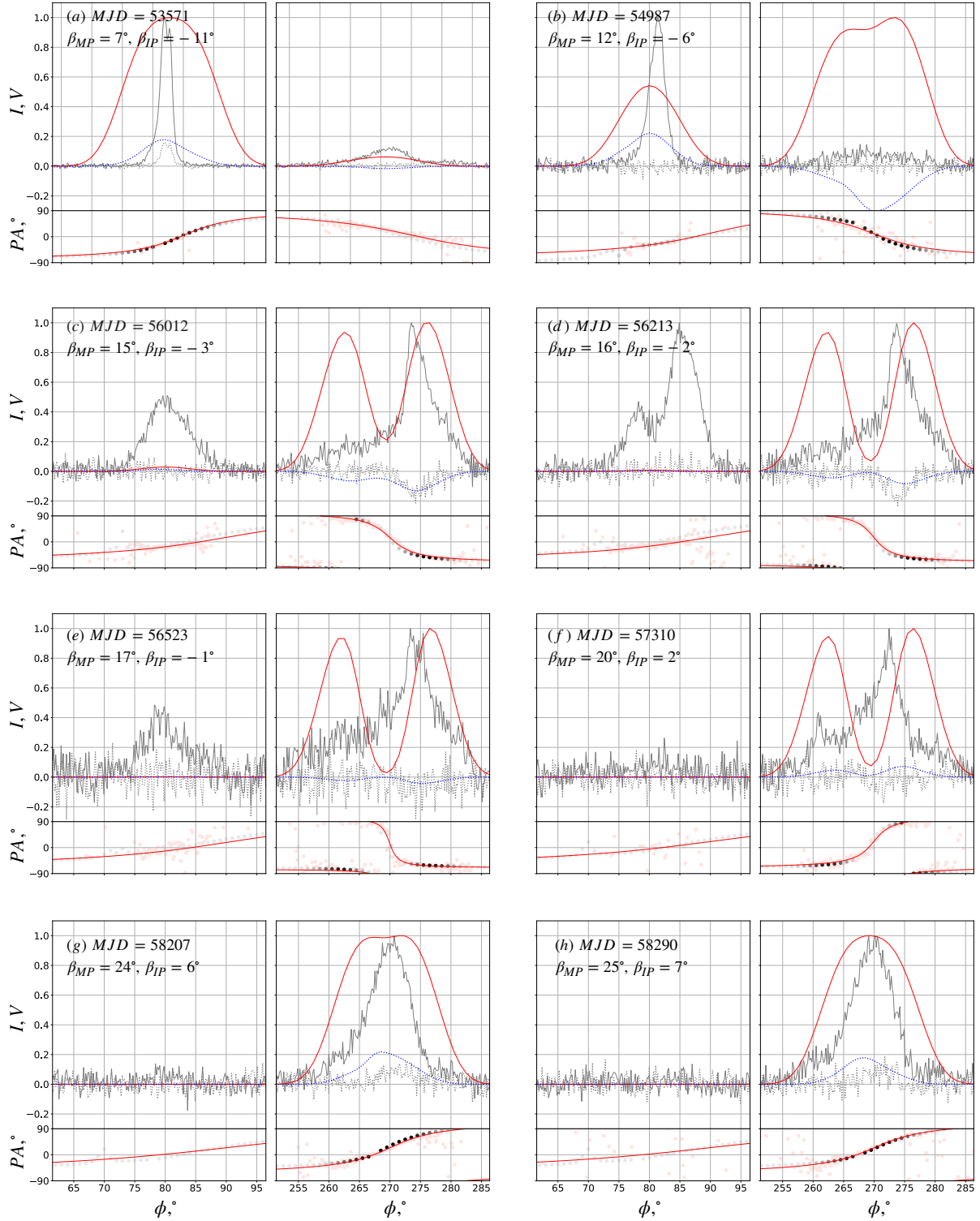}
    \caption{Simulations and observations of the radio intensity and polarization of PSR J1906+0746 as a function of pulsar phase, $\phi$, for different observational epochs. In each panel (a)-(h) of the Figure, the left sub-panel corresponds to the MP, and the right sub-panel corresponds to the IP. In each of the sub-panels, the top plots show the observed emission intensity (solid gray line) and circular polarization (dashed gray line), while the simulated intensity and circular polarization are shown as solid red and dashed blue lines, respectively. The bottom plots of each sub-panel show observed values of the position angle curve (red dots), RVM fits performed by \citet{Desvignes} (red solid line), and simulated values of the position angle (black dots).}
    \label{fig:pics}
\end{figure*}

Observations of pulsar PSR J1906+0746 at different epochs are shown in Figure~\ref{fig:pics}. In all panels, observed radio intensity is shown by a black solid line, and circle polarization is represented by a black dashed line. Being a nearly orthogonal rotator, this pulsar shows two pulses during its period of rotation. The first detected single pulse (on the left) is called the main pulse and the second (on the right) is called the interpulse (hereafter denoted MP and IP, respectively). In the bottom portion of each panel, observed position angle is shown by light red dots and a solid red line demonstrates the fitting with the RVM~\citep{Desvignes}.

As mentioned above, relativistic precession leads to significant changes in the shape of the mean pulse profile. First observations of the pulsar PSR J1906+0746 in 1998 show only the main pulse~\citep{PMPS}. Continuous observations were started in 2004~\citep{Lorimer}, when the  IP also started to appear (Figure~\ref{fig:pics}a). At that time, both the IP and the MP had single-peaked mean profiles. Their peaks were separated by half of a period ($180^{\circ}$ in pulsar phase), which is representative of an orthogonal rotator. The intensity of the MP was higher than the intensity of the IP (Figure~\ref{fig:pics}a). Circular polarization of the MP was positive, as was the derivative ${\rm d}PA/{\rm d} \phi$ of the position angle along the profile. Since the values $V$ and ${\rm d}PA/{\rm d} \phi$ had  the same sign, we can conclude that the MP corresponds to the X-mode, according to the criteria of ~\citet{PaperI} (see also the discussion in Section~\ref{sec:changeover}).

Over time, the IP became stronger, which enabled the recognition of its circular polarization at MJD 56012 (Figure~\ref{fig:pics}c). Since the IP had negative circular polarization $V$ and also negative derivative  ${\rm d}PA/{\rm d} \phi$, one can conclude that the IP also corresponds to the X-mode. 

At MJD $\sim 5600$, both pulses were roughly equally strong (Figure~\ref{fig:pics}c-d). Since the observed intensity peak of the IP was noticeably offset with respect to the maximum of ${\rm d} PA/{\rm d} \phi$, we assume that the IP was emitted as double-peaked (see also a hint of the second peak in Figure~\ref{fig:pics}f). The correlation of the signs of $V$ and ${\rm d}PA/{\rm d} \phi$ when $V$ is clearly identifiable is positive for both the MP and the IP, which correspond to the X-mode.

After MJD 56213 (Figure~\ref{fig:pics}d), the intensity of the MP decreased, while the sign of ${\rm d}PA/{\rm d} \phi$ remained the same (Figure~\ref{fig:pics}e). The IP was clearly visible, but its circular polarization was impossible to determine. Gradually, the derivative ${\rm d}PA/{\rm d} \phi$ in the IP became steeper. At some moment between MJD 56523 and MJD 57310, the sign of the derivative ${\rm d}PA/{\rm d} \phi$ changed in the IP, which implied the crossing of the magnetic axis at the polar cap, according to the prediction of the RVM (Figure~\ref{fig:pics}f). Unfortunately, at this time it was hard to clearly recognize the sign of the circular polarization in the IP (as with MP, which became almost invisible). However, the IP demonstrates a clearly positive circular polarization, $V$, in later observations. Also, after the crossing of the magnetic axis, the position angle derivative ${\rm d}PA/{\rm d} \phi$ remained positive in the IP (Figure~\ref{fig:pics}g-h). Thus, in agreement with theoretical expectations (see the discussion in Section~\ref{sec:changeover}), the quantities $V$ and ${\rm d}PA/{\rm d} \phi$ simultaneously changed their signs when the line of sight crossed the magnetic axis.  

\section{Theory and Simulations}
\label{sec:t&s}

\subsection{Basic equations}
Let us recall the key points related to the propagation theory of radio waves in the pulsar magnetosphere ~\citep{PaperI, PaperII}. First, as is well-known~\citep{L&GS, L&K}, there are two orthogonal modes which propagate in the pulsar magnetosphere: the extraordinary X-mode and the ordinary O-mode. The X-mode propagates along a straight line, while the O-mode experiences refraction, which takes place at small distances $r\leq r_{\rm O}$ from the star, where
\begin{equation}
    r_{\rm O} \sim 10^2 R \lambda_{4}^{1/3} \gamma_{100}^{1/3} B_{12}^{1/3} \nu_{\rm GHz}^{-2/3} P^{-1/5}.
    \label{eq:ro}
\end{equation}
Here, $R$ is the stellar radius, $\lambda = n_{\rm e}/n_{\rm GJ}$ is the pair creation multiplicity where $n_{\rm GJ} = \Omega B/(2 \pi ce)$ is the GJ number density, $\lambda_4 = \lambda / 10^4$, $\gamma_{100}$ is the normalized Lorentz-factor of secondary plasma ($\gamma_{100} = \gamma / 100$), $B_{12}$ is the normalized magnetic field on the surface ($B_{12} = B/10^{12}$ G), $\nu_{\rm GHz}$ is the frequency in GHz and $P$ is the period of rotation measured in seconds. The difference in propagation between two orthogonal modes was first taken into account by~\citet{B&A} and later discussed by various groups~\citep{BGI88, P&L1, P&L2, PaperI}. 

The second important propagation effect that has to be addressed is the limiting polarization effect (see, e.g.,~\citealt{Zheleznyakov}). Its importance for radio pulsars was first noted by~\citet{C&R} and later discussed in detail by~\citet{lim1, P&L2, lim3, andrianovbeskin2010, wanglaihan2010}. This effect arises because the plasma number density, $n_{\rm e}$, rapidly decreases with the distance $r$ from the star. In the region of a dense plasma, close to the stellar surface, the polarization characteristics of the wave depend on the direction of the external magnetic field. However, polarization characteristics almost freeze when the radiation propagates into the region of low plasma density. According to~\citet{C&R}, the freezing distance $r_{\rm esc}$ can be calculated as
\begin{equation}
    r_{\rm esc} \sim 10^3 R \lambda_{4}^{2/5} \gamma_{100}^{-6/5} B_{12}^{2/5} \nu_{\rm GHz}^{-2/5} P^{-1/5}.
    \label{eq:resc}
\end{equation}

As was noted by~\citet{andrianovbeskin2010, PaperI}, the most convenient method for studying the liming polarization effect for radio pulsars is the~\citet{KO} approach, which directly describes the evolution of the complex angle \mbox{$\Theta = \Theta_1 + i \Theta_2$} along the propagation of the radio wave: 
\begin{eqnarray}
\frac{{\rm d}\Theta_1}{{\rm d}l} = &&
\frac{\omega}{2c}{\rm Im}[\varepsilon_{xy}]
\nonumber \\
&& -\frac{1}{2}\frac{\omega}{c} \Lambda\cos[2\Theta_1-2\beta(l)-2\delta(l)]{\rm sinh}2\Theta_2,
\label{t1}\\
\frac{{\rm d}\Theta_2}{{\rm d}l} = &&
\frac{1}{2}\frac{\omega}{c}\Lambda\sin[2\Theta_1 - 2\beta(l)-2\delta(l)] {\rm cosh}2\Theta_2,
\label{t2}
\end{eqnarray}
where
\begin{equation}
\Lambda=\mp\sqrt{({\rm Re}[\varepsilon_{xy}])^2+\left(\frac{\varepsilon_{xx}-\varepsilon_{yy}}{2}\right)^2}.
\end{equation}
Here, $\Theta_1$ is the position angle of linear polarization (i.e., $\Theta_1 = PA$), $\Theta_2$ describes the circular polarization: \mbox{$\Theta_2 = 1/2 \tanh^{-1}(V/I)$} (here, $I$ and $V$ are two Stokes parameters), and $\varepsilon_{ik}$ is the dielectric tensor of the magnetized pair plasma. 

In homogeneous media, the normal modes of the plasma are described as $PA = \beta_B + \delta$ and $PA = \beta_B + \delta + \pi/2$ for the O-mode and X-mode, respectively. Here, the angle $\beta_B$ describes the orientation of the external magnetic field in the plane perpendicular to the wave vector $\mathbf {k}$. The nonzero angle $\delta$ results from the drift motion of plasma particles ${\bf U}/c = [{\bf E} \times {\bf B}]/{\bf B}^2$  in the corotational electric field $E \approx (\Omega r/c)B$:
\begin{equation}
\tan \delta = -\frac{\cos\theta_{\rm b}U_{x}/c}{\sin\theta_{\rm b} - U_y/c},    
\end{equation}
where $\theta_{\rm b}$ is the angle between the wave vector {\bf k} and the external magnetic field ${\bf B}$, the $x$-$y$ plane is perpendicular to the external magnetic field ${\bf B}$, and the unit vector in the $x$ direction is oriented in the ${\bf k}$-${\bf B}$ plane.

\subsection{Two orthogonal modes and polarization}\label{sec:changeover}

The main result of the propagation theory is the relation between the circular polarization $V$ and the derivative ${\rm d}(\beta_B + \delta)/{\rm d}r$ along the ray~\citep{PaperI}
\begin{equation}
    \frac{V}{I} \propto \frac{{\rm d}(\beta_B + \delta)/{\rm d}r}{\cos(2\Theta_1 - 2\beta_B - 2\delta)}.
    \label{eq:VI}
\end{equation}
This relation is valid when the shear of the external magnetic field along the ray trajectory is sufficiently large and when the sign of the circular polarization $V$ is determined not by the non-diagonal components of the dielectric tensor $\varepsilon_{ij}$ but by the sign of the derivative ${\rm d}(\beta_B + \delta)/{\rm d}r$ at the escape radius $r \approx r_{\rm esc}$. Moreover, as was shown by~\citet{wanglaihan2010}, for the dipolar geometry of the magnetic field, the derivative ${\rm d}\beta_B/{\rm d}r$ is proportional to the observable value ${\rm d} PA / {\rm d}\phi$ with the negative sign. Thus, equation~\ref{eq:VI} allows us to determine which of the two modes, ordinary or extraordinary, forms the mean profile of the radio emission. 

As was shown by~\citet{PaperII}, for most pulsars the freezing radius $r_{\rm esc}$ is sufficiently large that the derivative ${\rm d}\delta/{\rm d}r$ (whose sign is the opposite of the sign of ${\rm d}\beta_B/{\rm d}r$) plays the leading role. In this case,  the signs of ${\rm d} PA / {\rm d}\phi$ and $V$ are the same for the X-mode, and are opposite for the O-mode. At present, there is direct observational evidence that supports this correlation. Indeed, the main difference between the modes is that the O-mode experiences refraction, while the X-mode propagates along a straight line. Therefore, the widths of the mean profiles formed by the O-mode should be statistically larger than those of the X-mode profiles, which is indeed what has been observed~\citep{Beskin18}.

\begin{figure}
    \includegraphics[width=\columnwidth]{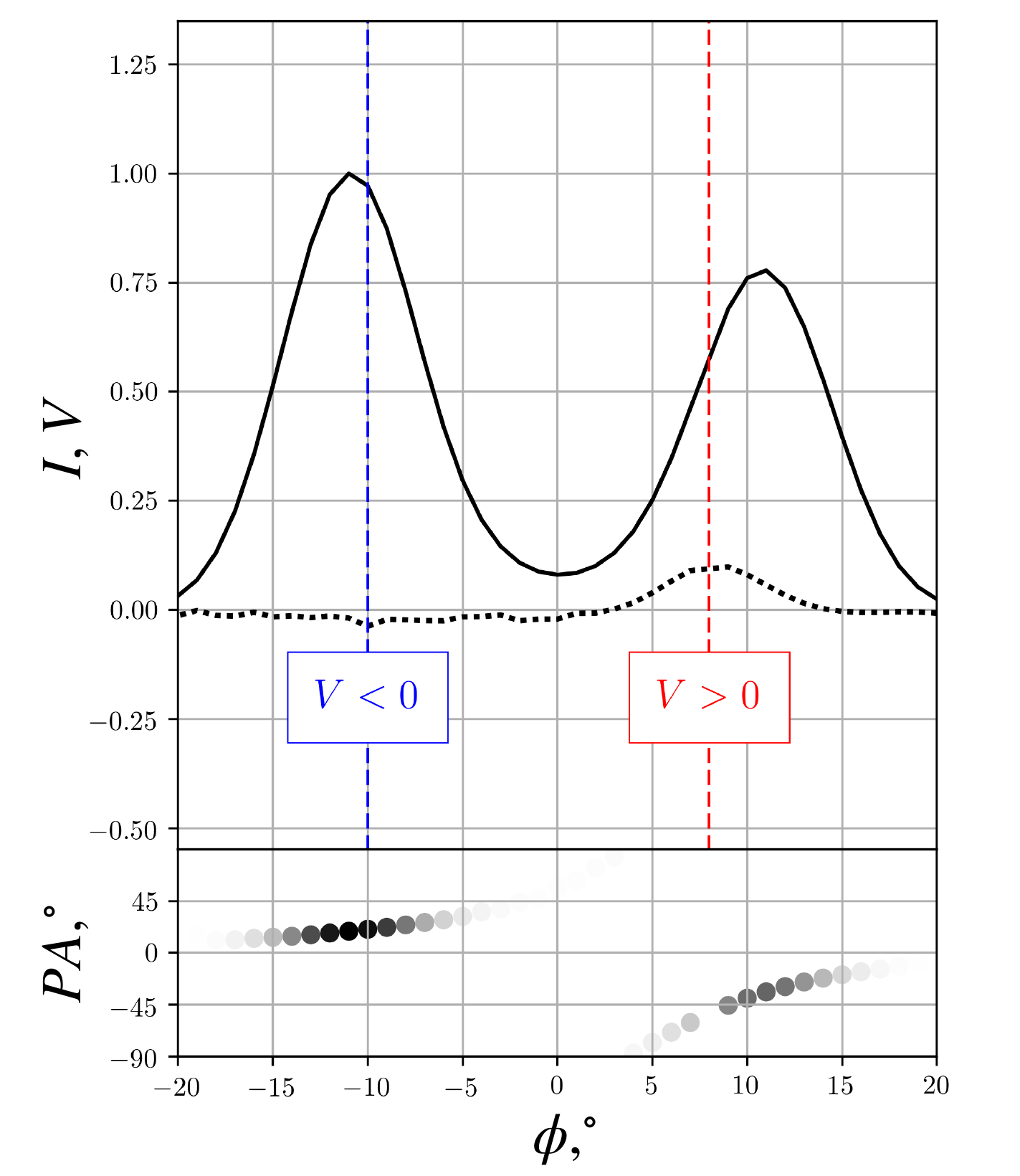}
    \caption{Simulated profile for a pulsar with parameters shown in Table~\ref{tab:par1}. Circular polarization $V$ has different signs across the profile, because polarization is formed at different heights.}
    \label{fig:V}
\end{figure}

However, there are examples of radio profiles where the correlation described above is not observed, i.e., there are observations for which the sign of the circular polarization $V$ along the mean profile varies, while the position angle curve $PA(\phi)$ does not show any transition to a different polarization mode. As was shown by~\citet{PaperII}, such behavior can  be also described in the framework of the propagation theory. Indeed, for small enough distances $r$ from the neutron star, one can neglect the angle $\delta$ in the sum $\beta_B + \delta$ because $\beta_B \gg \delta$. The fact that the derivatives ${\rm d}\delta/{\rm d}x$ and ${\rm d}\beta_B/{\rm d}x$ have different signs results in different signs of the circular polarization $V$ at different phases $\phi$ of the mean profile if the freezing radius $r_{esc}$ is located in the regions of different asymptotic behavior of the sum $\beta_B + \delta$. This means that the circular polarization $V$ is formed at different heights at different phases of the mean profile. This effect is quantitatively demonstrated in Figure~\ref{fig:V}, where we show a simulated profile for the propagation of the O-mode in the pulsar magnetosphere with parameters shown in Table~\ref{tab:par1}. 

\begin{table}
    \centering
    \caption{Parameters of the simulations shown in Figure~\ref{fig:V}. Here, $B_{12}=B/10^{12}$ is the normalized strength of the magnetic field near the pulsar surface, $P$ is the period of rotation in seconds, $\nu_{GHz}$ is the radiation frequency in GHz, $\lambda$ is the pair multiplicity, $\gamma$ is the Lorentz-factor of the secondary plasma, $\alpha$ is the inclination angle, $r_{em}$ is the emission height measured in units of the pulsar radius, and $f_0$ is a parameter used for determination of the magnetospheric pair plasma density in equation~\ref{eq:DENS}.}
    \label{tab:par1}
    \begin{tabular}{||c c c c c c c c c c||} 
        \hline
        $B_{12}$ & $P$ & $\nu_{\rm GHz}$ & $\lambda$ & $\gamma$ & $\alpha$ & $\beta$ & $r_{\rm em}$ & $f_0$\\ [0.5ex] 
         \hline\hline
         1.0 & 0.9 & 2.0 & 1000 & 250 & $45^o$ & $4^o$ & 30 & 0.7 \\  [1ex] 
        \hline
    \end{tabular}
\end{table}

From the usual correlation of the signs of $V$ and ${\rm d}PA/{\rm d} \phi$ one additional effect, which is specific to interpulse pulsars, can be noticed. As one can see from the RVM formula for the position angle~\citep{R&Cooke}
\begin{equation}
    PA = \arctan \Bigg( \frac{\sin \alpha \sin \phi}{\sin \alpha \cos (\alpha+\beta) \cos \phi - \sin (\alpha+\beta) \cos \alpha} \Bigg),
\end{equation}
during the crossing of the magnetic axis ($\beta \approx 0$), the derivative ${\rm d} PA / {\rm d}\phi$ changes its sign. As was mentioned above, the signs of $V$ and ${\rm d} PA / {\rm d}\phi$ are usually correlated for both modes. Therefore, the sign of $V$ also has to change to an opposite one after crossing the magnetic axis. This has been observed in PSR J0916+0746 at MJD $\sim 57000$. We also note that for almost all nearly orthogonal pulsars, the sign of the circular polarization $V$ remains constant when the position angle curve $PA(\phi)$ indicates that the mean profile is formed by one specific mode~\citep{JohnstonKramer}.

\subsection{Polar cap structure}\label{sec:cap}

\begin{figure*}
    \centering
    \includegraphics[width=\textwidth]{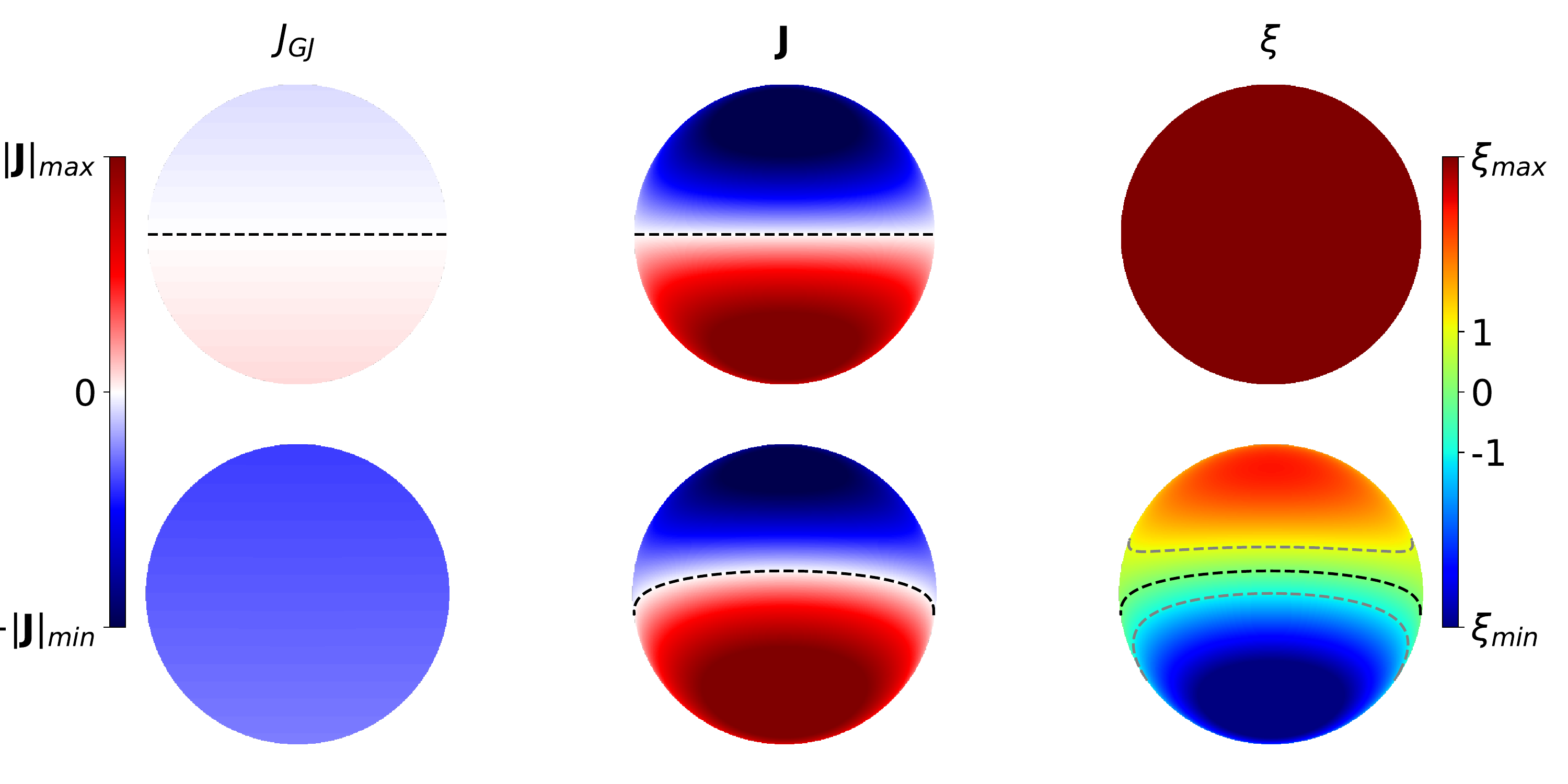}
    \caption{Distributions of the GJ current, $j_{GJ}$, electric current density, $j_{\parallel}$, and their ratio, $\xi = j_{\parallel}/j_{GJ}$ in the polar cap. The top row corresponds to the orthogonal rotator, $\alpha = 90^{\circ}$, and the bottom row corresponds to the case of PSR J1906+0746, $\alpha=81^{\circ}$.  A black dashed line accounts for the zero-value in all plots. Gray dashed lines mark the contours of $\xi=\pm1$ for $\alpha=81^{\circ}$, which bound the regions that can support pair production, $\xi<0$ and $\xi>1$. The plot of $\xi$ for the orthogonal rotator is saturated, since $\xi\gg 1$ and the whole polar cap supports active pair production in this case. These plots use analytical fits to the results of force-free simulations of pulsar magnetospheres by \citealt{Gralla}.}
    \label{fig:cap}
\end{figure*}

As was already mentioned, almost orthogonal pulsars ($\alpha \approx 90^{\circ}$), require separate consideration to determine the regions of plasma generation within the polar cap. As has been shown both analytically~\citep{Shibata, Beloborodov} and numerically~\citep{Timokhin, TimokhinArons}, under the assumption of free particle ejection from the surface of a neutron star, pair production only occurs in the regions that support a super-GJ current flow, $\xi=j_{\parallel}/j_{\rm GJ}>1$, where $j_{\rm GJ}$ is given by relation (\ref{GJ}), or in the regions with a magnetospheric return current $\xi<0$. For almost orthogonal rotators, numerical simulations of magnetospheres produce very high current density, $\xi \gg  1$, in  the entire polar cap ~\citep{BaiSpitkovsky}. In Figure \ref{fig:cap} we show the distributions of $j_{\rm GJ}$, $j_{\parallel}$ and $\xi$ at the polar cap of the orthogonal and the $81^{\circ}$ -inclined rotators, following analytical fits to simulation results by~\citet{Gralla}. As is clear from the plots, the region in which $0<\xi<1$, which does not allow for pair production, is small and localized near the magnetic axis.

For this reason, in the simulations described below we choose the number density of outgoing plasma, $n_{\rm e}$, to be:
\begin{equation}
     n_{\rm e} = 
     \lambda g(f) n_{\rm GJ}\,\zeta,  \qquad 
     g(f) = \frac{f^{2.5} {\rm exp}(-f^2)}{f^{2.5} + f_0^{2.5}}.
    \label{eq:DENS}
\end{equation}
Here $f = r_{\bot}^2/R_0^2$ is the characteristic polar cap area, $R_0 = (\Omega R/c)^{1/2}R$ is the characteristic polar cap radius, and $f_{0}$ models the diminishing of the number density $n_{\rm e}$ in the center (for $f_{0} = 0$ we have no 'hollow cone' at all). This correction mimics the suppression of the curvature radiation of accelerated particles and the associated pair production process near the center of a polar cap, where the curvature radius of magnetic field lines is large. Finally, $\zeta$ is a step function which defines the region of the polar cap that supports pair production, given by the condition $\xi<0$ or $\xi>1$.

\subsection{Simulation Results}

In this section we present the results of our modeling of the radio polarization of PSR J1906+0746 obtained by solving equations (\ref{t1})--(\ref{t2}) numerically \citep{PaperI}. We use realistic distributions of the number density of the relativistic pair plasma outflowing along open magnetic field lines (see Section \ref{sec:cap}). Also, we assume intensity of radio emission in the emission zone to be proportional to plasma density, as was specified before. Since the signs of the Stokes parameter $V$ and the derivative ${\rm d}PA/{\rm d}\phi$ are the same for both the MP and IP, we determine that the mean profile is formed by the X-mode. Also, as the pulsar is precessing, we calculate the evolution of the mean profiles for the X-mode as the impact angle $\beta$ changes with time. Our best model parameters for our simulation are shown in Table~\ref{tab:simpar}. Below, we discuss how polarization signatures change for different values of the plasma parameters $\lambda$ and $\gamma$. We use values of angles $\alpha$ and $\beta$ as inferred by \citet{Desvignes}. This significantly simplifies the problem and allows us to place constraints on the parameters of the magnetospheric pair plasma. The emission height $r_{\rm em}\approx 10R$ is set by the width of the profile and is also fixed in our calculations. 

\begin{table}
    \centering
    \caption{Parameters of the simulations shown in Figure~\ref{fig:pics}. Definitions are the same as in Table~\ref{tab:par1}.}
    \label{tab:simpar}
    \begin{tabular}{||c c c c c c c c||} 
        \hline
        $B_{12}$ & $P$ & $\nu_{\rm GHz}$ & $\lambda$ & $\gamma$ & $\alpha$ & $r_{\rm em}$ & $f_0$\\ [0.5ex] 
         \hline\hline
         1.0 & 0.14 & 1.38 & 1000 & 300 & $81^o$ & 10 & 0.2 \\  [1ex] 
        \hline
    \end{tabular}
\end{table}

A comparison of our simulations with observable profiles at different epochs is shown in Figure~\ref{fig:pics}. At MJD 53571, the intensity of the MP is higher than the intensity of the IP, and both their ratio and their single-peaked shape are well reproduced by numerical simulations (Figure~\ref{fig:pics}a). For the MP, both $V$ and ${\rm d} PA/{\rm d} \phi$ are positive; i.e., the correlation of their signs is also reproduced. At this and all subsequent epochs, the shapes of $PA$ for both the MP and the IP closely follows the observed ones, as well as the RVM prediction. Furthermore, the degree of circular polarization is well reproduced by our calculations. However, while the single shape for both pulses is reproduced, our results show that the MP is significantly wider compared to what is observed. 

The following epoch (Figure~\ref{fig:pics}b) illustrates the deviation of the ratio of the intensities of the MP and IP. Although the $I$ and $V$ shapes are not reproduced numerically, the shape of $PA$ is in  good agreement with the observations for both pulses. Also, while the simulated $V$ and ${\rm d} PA/{\rm d} \phi$ have the same signs along both pulses, as required for the X-mode, there is no clearly detected circular polarization at this epoch. 

At MJD 56012, the simulated MP becomes significantly weaker than the IP (Figure~\ref{fig:pics}c). It stays invisible throughout profiles at later epochs in simulations, while it is still visible in observations until MJD $\sim 57000$. The IP is double-peaked, with the left hump being slightly weaker than the right one due to the synchrotron absorption that happens at large distances in the magnetosphere. Simulations reproduce the observed shapes of $V$ and ${\rm d} PA/{\rm d} \phi$ as well as the degree of $V$. The correlation of signs is also in agreement with observations: $V$ and ${\rm d} PA/{\rm d} \phi$ are both negative for the IP. Again, while the simulated IP reproduces the double-peaked shape, there is a significant difference from the observed intensity shape. We find that for higher values of $\lambda$, the resulting strong synchrotron absorption makes the IP intensity profile more compatible with observations; however, the $PA$ curve becomes significantly distorted in this case. Similar conclusions hold until MJD $\sim 57000$ (panels d and e in Figure~\ref{fig:pics}).

As the line of sight approaches the magnetic axis in the interpulse, $\beta_{\rm IP}=0$, the $PA(\phi)$ becomes steeper. After the crossing (at the moment in time between panels e and f), both $V$ and ${\rm d}PA/{\rm d} \phi$ change their signs, in agreement with the prediction of the propagation theory. Finally, Figures~\ref{fig:pics}g-h illustrate the latest available observations at MJD $\sim 58000$. The MP is again invisible, while the IP shows roughly the same behavior as MP at comparable impact angles, $\beta_{\rm IP} \approx \beta_{\rm MP}$ (see Figure~\ref{fig:pics}a). At these late epochs, we reproduce both the single-peak shape of the IP and the behavior of both $PA(\phi)$ and $V$. The simulated profile is slightly wider than the observed one.

To summarize, the choice of plasma parameters $\lambda\sim10^3$ and $\gamma\sim 300$ allows to reproduce both the position angle profile and the degree and sign of the circular polarization in observations (a poor fit of circular polarization at MJD 54987, panel b, is mainly due to a poor fit of the intensity, since $V=I \tanh(2\Theta_2)$). Higher values of $\lambda$ or smaller values of $\gamma$ result in stronger propagation effects, and thus a higher degree of circular polarization and more significant deviations of the $PA$ profile from the RVM prediction. In contrast, smaller $\lambda$ and higher $\gamma$ lead to a lower degree of circular polarization, $V$, or a change in the sign of $V$ along the same profile, as described in Sec.~\ref{sec:changeover} (see also Figure \ref{fig:V}). As discussed above, higher values of $\lambda$ may better describe the weakness of the first peak of the IP at MJD 56000-57000. However, given the high uncertainty in the initial intensity profile, we decided to focus on fitting the polarization signatures. 

Finally, we note that our simple model for the intensity works especially poorly for the MP (as well as at the boundary of the IP beam, when it just starts to appear). In this regard, we would like to point to an interesting circumstance. As we have seen above, due to the precession of the rotation axis, the observer's line of sight crosses the stellar equator ($\beta = 90^{\circ}-\alpha$, or $\beta_{IP}=-\beta_{MP}$; the moment of time between panels a and b in Figure \ref{fig:pics}) at epoch MJD$\sim 54400$. If the magnetic field at the neutron star surface would correspond to the field of a centered dipole, then the MP and IP would have the same shape at this time, for any model of the initial intensity (see also the bottom panel in Figure \ref{fig:crossing}). However, observations show quite different radio intensity profiles. Therefore, this fact can serve as an indication that the magnetic field at the stellar surface differs from the dipolar one for the pulsar PSR J1906+0746. According to~\citet{JohnstonKramer}, the same is true for the pulsar PSR J1828-1101, since the condition $\beta_{IP}=-\beta_{MP}$ is also fulfilled. In the case of PSR J1906+0746, the intensity profiles of the MP and the IP also differ significantly.

\section{Discussion}
\label{sec:conslusion}

In this paper, we present simulations of the formation of the radio polarization of the nearly orthogonal pulsar PSR J1906+0746 using the method developed by \citet{PaperI}. Our numerical models have been able to explain the general shape, sign and degree of the circular polarization, $V$, for many observational epochs. In particular, the change of sign of $V$ at the moment when the line of sight crosses the IP magnetic axis is a very robust prediction of the propagation theory. We find that the observed circular polarization is best explained for a multiplicity of secondary plasma of $\sim 10^3$, and a Lorentz factor of $\sim$ a few hundred. For these parameters of the pulsar plasma, the profile of the position angle is not dramatically distorted and matches with the RVM fits shown in \citet{Desvignes}. 

While our modeling shows that the radio polarization can be robustly modeled, it also shows that the assumptions we used for describing the shape of the radio beam are still overly simplistic. This is particularly evident when we  model the MP and the edge of the IP beam. The observed behavior of the MP is even more complicated, and potentially requires a non-dipolar surface magnetic field. Analytical models of the near-field structure of non-dipolar magnetospheres exist \citep{Gralla} and have been used to explain the puzzling behavior of the X-ray light curves of millisecond pulsars \citep{Lockhart, Bilous2019}. Despite this, we think the main uncertainty lies in our assumption that the radio intensity is proportional to the density of the secondary plasma. For example, recent particle-in-cell numerical investigations of the pulsar emission mechanism suggest the important role of the non-uniformity of the plasma flow close to the polar cap, bringing in extra dependencies \citep{Philippov2020}. An additional complication arises because the mean profile is formed by the X-mode, which cannot be directly excited by the plasma currents and must be a "propagation effect" by itself\footnote{This potentially explains why we require a somewhat high emission altitude, $10R$, above the surface.}. This may also explain why we require a somewhat small multiplicity value, compared to traditional estimates of $\sim 10^5$ \citep{TimokhinHarding2015}. In fact, polarization may be freezing at lower altitude if the emission escapes through the local under-dense plasma zone, which can be seen as effectively lower density of a uniform flow. As these theoretical models and ideas are developed further, modeling of the light curves of PSR J1906+0746 may provide key tests of the theory.

To conclude, we want to emphasize the uniqueness of this pulsar for future tests of theories of the pulsar radio emission mechanism. For example, the fact that the IP is observed both above and below the magnetic axis rules out theories that rely on the presence of ions in the plasma flow \citep{Ruderman1980,Jones2014}, at least as a generic model applicable to all pulsars. This is because ion extraction is only possible in the half of the polar cap that supports a volume return current (the lower half of the polar cap in Figure \ref{fig:cap}). These observations also rule out early models of stationary pair cascades, which did not take the effects of general relativity into account and predicted that pair production, and thus radio emission, would be active only in one half of the polar cap \citep{Arons1979}. PSR J1906+0746 emphasizes the role of unique, long-term observations, which are sometimes more useful to constraint theories than the continuous increase of observed sources.

\section{Acknowledgements} The authors thank G. Desvignes and M. Kramer for the possibility to work with observational data, and A. Jessner and M. Kramer for fruitful discussions. This work was partially supported by the government of the Russian Federation (agreement No. 05.Y09.21.0018) and by the Russian Foundation for Basic Research (Grant no. 20-02-00469). Research at the Flatiron Institute is supported by the Simons Foundation.

\section{Data availability}
The observational data analyzed in this article were provided by Greg Desvignes (\href{mailto:gdesvignes.astro@gmail.com}{gdesvignes.astro@gmail.com}) by permission. The simulation data will be shared on request to the corresponding author, Alisa Galishnikova.

\bibliographystyle{mnras}
\bibliography{bibliography}

\begin{thebibliography}{}
\makeatletter
\relax
\def\mn@urlcharsother{\let\do\@makeother \do\$\do\&\do\#\do\^\do\_\do\%\do\~}
\def\mn@doi{\begingroup\mn@urlcharsother \@ifnextchar [ {\mn@doi@}
  {\mn@doi@[]}}
\def\mn@doi@[#1]#2{\def\@tempa{#1}\ifx\@tempa\@empty \href
  {http://dx.doi.org/#2} {doi:#2}\else \href {http://dx.doi.org/#2} {#1}\fi
  \endgroup}
\def\mn@eprint#1#2{\mn@eprint@#1:#2::\@nil}
\def\mn@eprint@arXiv#1{\href {http://arxiv.org/abs/#1} {{\tt arXiv:#1}}}
\def\mn@eprint@dblp#1{\href {http://dblp.uni-trier.de/rec/bibtex/#1.xml}
  {dblp:#1}}
\def\mn@eprint@#1:#2:#3:#4\@nil{\def\@tempa {#1}\def\@tempb {#2}\def\@tempc
  {#3}\ifx \@tempc \@empty \let \@tempc \@tempb \let \@tempb \@tempa \fi \ifx
  \@tempb \@empty \def\@tempb {arXiv}\fi \@ifundefined
  {mn@eprint@\@tempb}{\@tempb:\@tempc}{\expandafter \expandafter \csname
  mn@eprint@\@tempb\endcsname \expandafter{\@tempc}}}

\bibitem[\protect\citeauthoryear{{Andrianov} \& {Beskin}}{{Andrianov} \&
  {Beskin}}{2010}]{andrianovbeskin2010}
{Andrianov} A.~S.,  {Beskin} V.~S.,  2010, \mn@doi [Astron. Lett.]
  {10.1134/S1063773710040031}, \href
  {http://adsabs.harvard.edu/abs/2010AstL...36..248A} {36, 248}

\bibitem[\protect\citeauthoryear{{Arons} \& {Scharlemann}}{{Arons} \&
  {Scharlemann}}{1979}]{Arons1979}
{Arons} J.,  {Scharlemann} E.~T.,  1979, \mn@doi [\apj] {10.1086/157250}, \href
  {https://ui.adsabs.harvard.edu/abs/1979ApJ...231..854A} {231, 854}

\bibitem[\protect\citeauthoryear{{Bai} \& {Spitkovsky}}{{Bai} \&
  {Spitkovsky}}{2010}]{BaiSpitkovsky}
{Bai} X.-N.,  {Spitkovsky} A.,  2010, \mn@doi [\apj]
  {10.1088/0004-637X/715/2/1282}, \href
  {https://ui.adsabs.harvard.edu/abs/2010ApJ...715.1282B} {715, 1282}

\bibitem[\protect\citeauthoryear{{Barnard}}{{Barnard}}{1986}]{lim1}
{Barnard} J.~J.,  1986, \mn@doi [\apj] {10.1086/164073}, \href
  {http://adsabs.harvard.edu/abs/1986ApJ...303..280B} {303, 280}

\bibitem[\protect\citeauthoryear{{Barnard} \& {Arons}}{{Barnard} \&
  {Arons}}{1986}]{B&A}
{Barnard} J.~J.,  {Arons} J.,  1986, \mn@doi [\apj] {10.1086/163979}, \href
  {http://adsabs.harvard.edu/abs/1986ApJ...302..138B} {302, 138}

\bibitem[\protect\citeauthoryear{{Beloborodov}}{{Beloborodov}}{2008}]{Beloborodov}
{Beloborodov} A.~M.,  2008, \mn@doi [\apj] {2008ApJ...683L..41B}, \href
  {https://ui.adsabs.harvard.edu/abs/1997MNRAS.287..262S/abstract} {683, L41}

\bibitem[\protect\citeauthoryear{{Beskin}}{{Beskin}}{2018}]{Beskin18}
{Beskin} V S.,  2018, \mn@doi [Phys. Uspekhi] {10.3367/UFNe.2017.10.038216},
  \href {https://ui.adsabs.harvard.edu/abs/2018PhyU...61..353B/abstract} {61,
  353}

\bibitem[\protect\citeauthoryear{{Beskin} \& {Philippov}}{{Beskin} \&
  {Philippov}}{2012}]{PaperI}
{Beskin} V.~S.,  {Philippov} A.~A.,  2012, \mn@doi [\mnras]
  {10.1111/j.1365-2966.2012.20988.x}, \href
  {https://ui.adsabs.harvard.edu/abs/2012MNRAS.425..814B} {425, 814}

\bibitem[\protect\citeauthoryear{{Beskin}, {Gurevich}  \& {Istomin}}{{Beskin}
  et~al.}{1988}]{BGI88}
{Beskin} V.~S.,  {Gurevich} A.~V.,   {Istomin} I.~N.,  1988, \mn@doi [\apss]
  {10.1007/BF00637577}, \href
  {http://adsabs.harvard.edu/abs/1988Ap\%26SS.146..205B} {146, 205}

\bibitem[\protect\citeauthoryear{{Bilous} et~al.,}{{Bilous}
  et~al.}{2019}]{Bilous2019}
{Bilous} A.~V.,  et~al., 2019, \mn@doi [\apjl] {10.3847/2041-8213/ab53e7},
  \href {https://ui.adsabs.harvard.edu/abs/2019ApJ...887L..23B} {887, L23}

\bibitem[\protect\citeauthoryear{{Cheng} \& {Ruderman}}{{Cheng} \&
  {Ruderman}}{1979}]{C&R}
{Cheng} A.~F.,  {Ruderman} M.~A.,  1979, \mn@doi [\apj] {10.1086/156959}, \href
  {http://adsabs.harvard.edu/abs/1979ApJ...229..348C} {229, 348}

\bibitem[\protect\citeauthoryear{{Cheng} \& {Ruderman}}{{Cheng} \&
  {Ruderman}}{1980}]{Ruderman1980}
{Cheng} A.~F.,  {Ruderman} M.~A.,  1980, \mn@doi [\apj] {10.1086/157661}, \href
  {https://ui.adsabs.harvard.edu/abs/1980ApJ...235..576C} {235, 576}

\bibitem[\protect\citeauthoryear{{Desvignes} et~al.,}{{Desvignes}
  et~al.}{2019}]{Desvignes}
{Desvignes} G.,  et~al., 2019, \mn@doi [Science] {10.1126/science.aav7272},
  \href {https://ui.adsabs.harvard.edu/abs/2019Sci...365.1013D} {365, 1013}

\bibitem[\protect\citeauthoryear{{Fussell}, {Luo}  \& {Melrose}}{{Fussell}
  et~al.}{2003}]{abs1}
{Fussell} D.,  {Luo} Q.,   {Melrose} D.~B.,  2003, \mn@doi [\mnras]
  {10.1046/j.1365-8711.2003.06767.x}, \href
  {http://adsabs.harvard.edu/abs/2003MNRAS.343.1248F} {343, 1248}

\bibitem[\protect\citeauthoryear{{Goldreich} \& {Julian}}{{Goldreich} \&
  {Julian}}{1969}]{GoldreichJulian}
{Goldreich} P.,  {Julian} W.~H.,  1969, \mn@doi [\apj] {10.1086/150119}, \href
  {https://ui.adsabs.harvard.edu/abs/1969ApJ...157..869G} {157, 869}

\bibitem[\protect\citeauthoryear{{Gralla}, {Lupsasca}  \& {Philippov}}{{Gralla}
  et~al.}{2017}]{Gralla}
{Gralla} S.~E.,  {Lupsasca} A.,   {Philippov} A.,  2017, \mn@doi [\apj]
  {10.3847/1538-4357/aa978d}, \href
  {https://ui.adsabs.harvard.edu/abs/2017ApJ...851..137G} {851, 137}

\bibitem[\protect\citeauthoryear{{Hakobyan}, {Beskin}  \&
  {Philippov}}{{Hakobyan} et~al.}{2017a}]{PaperII}
{Hakobyan} H.~L.,  {Beskin} V.~S.,   {Philippov} A.~A.,  2017a, \mn@doi
  [\mnras] {10.1093/mnras/stx1025}, \href
  {https://ui.adsabs.harvard.edu/abs/2017MNRAS.469.2704H} {469, 2704}

\bibitem[\protect\citeauthoryear{{Hakobyan}, {Philippov}, {Beskin}, {Novoselov}
   \& {Rashkovetskyi}}{{Hakobyan} et~al.}{2017b}]{Hakobyan}
{Hakobyan} H.~L.,  {Philippov} A.~A.,  {Beskin} V.~S.~{Galishnikova} A.~K.,
  {Novoselov} E.~M.,   {Rashkovetskyi} M.~M.,  2017b, \mn@doi [J. Phys. Conf.
  Ser.] {10.1088/1742-6596/932/1/012018}, \href
  {https://ui.adsabs.harvard.edu/abs/2017JPhCS.932a2018H/abstract} {932,
  012018}

\bibitem[\protect\citeauthoryear{{Hankins} \& {Rankin}}{{Hankins} \&
  {Rankin}}{2010}]{H&R}
{Hankins} T.~H.,  {Rankin} J.~M.,  2010, \mn@doi [\aj]
  {10.1088/0004-6256/139/1/168}, \href
  {http://adsabs.harvard.edu/abs/2010AJ....139..168H} {139, 168}

\bibitem[\protect\citeauthoryear{{Hewish}, {Bell}, {Pilkington}, {Scott}  \&
  {Collins}}{{Hewish} et~al.}{1968}]{Hewish}
{Hewish} A.,  {Bell} S.~J.,  {Pilkington} J.~D.~H.,  {Scott} P.~F.,   {Collins}
  R.~A.,  1968, \mn@doi [\nat] {10.1038/217709a0}, \href
  {http://adsabs.harvard.edu/abs/1968Natur.217..709H} {217, 709}

\bibitem[\protect\citeauthoryear{{Johnston} \& {Kerr}}{{Johnston} \&
  {Kerr}}{2016}]{J&K}
{Johnston} S.,  {Kerr} M.,  2016, \mn@doi [\mnras] {10.1093/mnras/stx3095},
  \href {http://adsabs.harvard.edu/abs/2018MNRAS.474.4629J} {474, 4629}

\bibitem[\protect\citeauthoryear{{Johnston} \& {Kramer}}{{Johnston} \&
  {Kramer}}{2019}]{JohnstonKramer}
{Johnston} S.,  {Kramer} M.,  2019, \mn@doi [\mnras] {10.1093/mnras/stz2865},
  \href {https://ui.adsabs.harvard.edu/abs/2019MNRAS.490.4565J/abstract} {490,
  4565}

\bibitem[\protect\citeauthoryear{{Jones}}{{Jones}}{2014}]{Jones2014}
{Jones} P.~B.,  2014, \mn@doi [\mnras] {10.1093/mnras/stu1916}, \href
  {https://ui.adsabs.harvard.edu/abs/2014MNRAS.445.2297J} {445, 2297}

\bibitem[\protect\citeauthoryear{{Kravtsov} \& {Orlov}}{{Kravtsov} \&
  {Orlov}}{1990}]{KO}
{Kravtsov} Y.~A.,  {Orlov} Y.~I.,  1990, {Geometrical Optics of Inhomogeneous
  Media}.
{Springer-Verlag}, Berlin

\bibitem[\protect\citeauthoryear{{Lockhart}, {Gralla}, {{\"O}zel}  \&
  {Psaltis}}{{Lockhart} et~al.}{2019}]{Lockhart}
{Lockhart} W.,  {Gralla} S.~E.,  {{\"O}zel} F.,   {Psaltis} D.,  2019, \mn@doi
  [\mnras] {10.1093/mnras/stz2524}, \href
  {https://ui.adsabs.harvard.edu/abs/2019MNRAS.490.1774L/abstract} {490, 1774}

\bibitem[\protect\citeauthoryear{{Lorimer} \& {Kramer}}{{Lorimer} \&
  {Kramer}}{2012}]{L&K}
{Lorimer} D.~R.,  {Kramer} M.,  2012, {Handbook of Pulsar Astronomy}.
{Cambridge University Press}, Cambridge

\bibitem[\protect\citeauthoryear{Lorimer et~al.,}{Lorimer
  et~al.}{2006}]{Lorimer}
Lorimer D.~R.,  et~al., 2006, \mn@doi [\apj] {10.1086/499918}, \href
  {https://ui.adsabs.harvard.edu/abs/2006ApJ...640..428L/abstract} {640, 428}

\bibitem[\protect\citeauthoryear{{Lyne} \& {Graham-Smith}}{{Lyne} \&
  {Graham-Smith}}{2012}]{L&GS}
{Lyne} A.,  {Graham-Smith} F.,  2012, {Pulsar Astronomy}.
{Cambridge University Press}, Cambridge

\bibitem[\protect\citeauthoryear{{Lyne} \& {Manchester}}{{Lyne} \&
  {Manchester}}{1988}]{L&M}
{Lyne} A.~G.,  {Manchester} R.~N.,  1988, \mn@doi [\mnras]
  {10.1093/mnras/234.3.477}, \href
  {http://adsabs.harvard.edu/abs/1988MNRAS.234..477L} {234, 477}

\bibitem[\protect\citeauthoryear{{Lyubarskii} \& {Petrova}}{{Lyubarskii} \&
  {Petrova}}{1998}]{P&L1}
{Lyubarskii} Y.~E.,  {Petrova} S.~A.,  1998, \aap, \href
  {http://adsabs.harvard.edu/abs/1998A\%26A...333..181L} {333, 181}

\bibitem[\protect\citeauthoryear{Manchester et~al.,}{Manchester
  et~al.}{2001}]{PMPS}
Manchester R.~N.,  et~al., 2001, \mn@doi [\mnras]
  {10.1046/j.1365-8711.2001.04751.x}, \href
  {https://ui.adsabs.harvard.edu/abs/2001MNRAS.328...17M/abstract} {328, 17}

\bibitem[\protect\citeauthoryear{{Melrose} \& {Luo}}{{Melrose} \&
  {Luo}}{2004}]{luomelrose2004}
{Melrose} D.~B.,  {Luo} Q.,  2004, \mn@doi [\mnras]
  {10.1111/j.1365-2966.2004.07986.x}, \href
  {http://adsabs.harvard.edu/abs/2004MNRAS.352..915M} {352, 915}

\bibitem[\protect\citeauthoryear{{Mikhailovskii}, {Onishchenko},
  {Suramlishvili}  \& {Sharapov}}{{Mikhailovskii}
  et~al.}{1982}]{mikhailovskii82}
{Mikhailovskii} A.~B.,  {Onishchenko} O.~G.,  {Suramlishvili} G.~I.,
  {Sharapov} S.~E.,  1982, Soviet Astronomy Letters, \href
  {http://adsabs.harvard.edu/abs/1982SvAL....8..369M} {8, 369}

\bibitem[\protect\citeauthoryear{{Oster} \& {Sieber}}{{Oster} \&
  {Sieber}}{1976}]{O&S}
{Oster} L.,  {Sieber} W.,  1976, \mn@doi [\apj] {10.1086/154820}, \href
  {https://ui.adsabs.harvard.edu/abs/1976ApJ...210..220O/abstract} {210, 220}

\bibitem[\protect\citeauthoryear{Perera et~al.,}{Perera et~al.}{2010}]{Perera}
Perera B. B.~P.,  et~al., 2010, \mn@doi [\apj] {10.1088/0004-637X/721/2/1193},
  \href {https://ui.adsabs.harvard.edu/abs/2010ApJ...721.1193P/abstract} {721,
  1193}

\bibitem[\protect\citeauthoryear{{Petrova}}{{Petrova}}{2006}]{lim3}
{Petrova} S.~A.,  2006, \mn@doi [\mnras] {10.1111/j.1365-2966.2006.10246.x},
  \href {http://adsabs.harvard.edu/abs/2006MNRAS.368.1764P} {368, 1764}

\bibitem[\protect\citeauthoryear{{Petrova} \& {Lyubarskii}}{{Petrova} \&
  {Lyubarskii}}{2000}]{P&L2}
{Petrova} S.~A.,  {Lyubarskii} Y.~E.,  2000, \aap, \href
  {http://adsabs.harvard.edu/abs/2000A\%26A...355.1168P} {355, 1168}

\bibitem[\protect\citeauthoryear{Philippov, Timokhin  \& Spitkovsky}{Philippov
  et~al.}{2020}]{Philippov2020}
Philippov A.,  Timokhin A.,   Spitkovsky A.,  2020, \mn@doi [Phys. Rev. Lett.]
  {10.1103/PhysRevLett.124.245101}, 124, 245101

\bibitem[\protect\citeauthoryear{{Radhakrishnan} \& {Cooke}}{{Radhakrishnan} \&
  {Cooke}}{1969}]{R&Cooke}
{Radhakrishnan} V.,  {Cooke} D.~J.,  1969, \aplett, \href
  {http://adsabs.harvard.edu/abs/1969ApL.....3..225R} {3, 225}

\bibitem[\protect\citeauthoryear{{Rankin}}{{Rankin}}{1993}]{Rankin}
{Rankin} J.,  1993, \mn@doi [\aj] {10.1086/172361}, \href
  {https://ui.adsabs.harvard.edu/abs/1993ApJ...405..285R/abstract} {405, 285}

\bibitem[\protect\citeauthoryear{{Shibata}}{{Shibata}}{1997}]{Shibata}
{Shibata} S.,  1997, \mn@doi [\mnras] {10.1093/mnras/287.2.262}, \href
  {https://ui.adsabs.harvard.edu/abs/1997MNRAS.287..262S/abstract} {287, 262}

\bibitem[\protect\citeauthoryear{{Tauris} \& {Manchester}}{{Tauris} \&
  {Manchester}}{1998}]{T&M}
{Tauris} T.~M.,  {Manchester} R.~N.,  1998, \mn@doi [\mnras]
  {10.1046/j.1365-8711.1998.01369.x}, \href
  {http://adsabs.harvard.edu/abs/1998MNRAS.298..625T} {298, 625}

\bibitem[\protect\citeauthoryear{{Timokhin}}{{Timokhin}}{2010}]{Timokhin}
{Timokhin} A.~N.,  2010, \mn@doi [\mnras] {10.1111/j.1365-2966.2010.17286.x},
  \href {https://ui.adsabs.harvard.edu/abs/2010MNRAS.408.2092T/abstract} {408,
  2092}

\bibitem[\protect\citeauthoryear{{Timokhin} \& {Arons}}{{Timokhin} \&
  {Arons}}{2013}]{TimokhinArons}
{Timokhin} A.~N.,  {Arons} J.,  2013, \mn@doi [\mnras] {10.1093/mnras/sts298},
  \href {https://ui.adsabs.harvard.edu/abs/2013MNRAS.429...20T} {429, 20}

\bibitem[\protect\citeauthoryear{{Timokhin} \& {Harding}}{{Timokhin} \&
  {Harding}}{2015}]{TimokhinHarding2015}
{Timokhin} A.~N.,  {Harding} A.~K.,  2015, \mn@doi [\apj]
  {10.1088/0004-637X/810/2/144}, \href
  {https://ui.adsabs.harvard.edu/abs/2015ApJ...810..144T} {810, 144}

\bibitem[\protect\citeauthoryear{Venkatraman~Krishnan, Bailes, van Straten,
  Keane, Kramer, Bhat, Flynn  \& Os{\l}owski}{Venkatraman~Krishnan
  et~al.}{2019}]{Krishnan}
Venkatraman~Krishnan V.,  Bailes M.,  van Straten W.,  Keane E.~F.,  Kramer M.,
   Bhat N. D.~R.,  Flynn C.,   Os{\l}owski S.,  2019, \mn@doi [\apj]
  {10.3847/2041-8213/ab0a03}, \href
  {https://ui.adsabs.harvard.edu/abs/2019ApJ...873L..15V/abstract} {873, L15}

\bibitem[\protect\citeauthoryear{{Wang}, {Lai}  \& {Han}}{{Wang}
  et~al.}{2010}]{wanglaihan2010}
{Wang} C.,  {Lai} D.,   {Han} J.,  2010, \mn@doi [\mnras]
  {10.1111/j.1365-2966.2009.16074.x}, \href
  {http://adsabs.harvard.edu/abs/2010MNRAS.403..569W} {403, 569}

\bibitem[\protect\citeauthoryear{{Weltevrede} \& {Johnston}}{{Weltevrede} \&
  {Johnston}}{2008}]{W&J}
{Weltevrede} P.,  {Johnston} S.,  2008, \mn@doi [\mnras]
  {10.1111/j.1365-2966.2008.13950.x}, \href
  {http://adsabs.harvard.edu/abs/2008MNRAS.391.1210W} {391, 1210}

\bibitem[\protect\citeauthoryear{{Zheleznyakov}}{{Zheleznyakov}}{1996}]{Zheleznyakov}
{Zheleznyakov} V.~V.,  1996, {Radiation in Astrophysical Plasmas}.
{Springer-Verlag}, Berlin

\makeatother
\end{thebibliography}

\label{lastpage}

\bsp	
\end{document}